# Parity-time symmetry breaking enables swarming motility in *Caenorhabditis elegans*

## Authors

Mustafa Basaran[1,2,3], Tevfik Can Yüce[2], Ali Keçebaş[4], Baha Altın[1], Yusuf Ilker Yaman[1,3], Esin Demir[2], Coşkun Kocabaş[5], Şahin K. Özdemir[5]†, and Aşkın Kocabaş[1,2,6,7]†

[1]Department of Physics, Koç University, Sarıyer, 34450 Istanbul, Turkey

[2]Bio-Medical Sciences and Engineering Program, Koç University, Sarıyer, 34450 Istanbul, Turkey

[3]Current adres: Sciences and Engineering Program, Harvard University, Cambridge 02138, MA

[4]Department of Engineering Science and Mechanics, The Pennsylvania State University, University Park, 16802, PA

[5]Department of Materials, University of Manchester, Manchester, M13 9PL, UK

[6]Koç University Surface Science and Technology Center, Koç University, Sarıyer, 34450 Istanbul, Turkey

[7]Koç University Research Center for Translational Medicine, Koç University, Sarıyer, 34450 Istanbul, Turkey

†Corresponding author: sko9@psu.edu , akocabas@ku.edu.tr

# Abstract

Nonreciprocal interactions break action-reaction symmetry in systems of interacting bodies. This process inevitably introduces non-Hermitian dynamics which with its hallmark signature called exceptional points (EPs) has been a subject of intense research across different disciplines ranging from photonics to metamaterials. Whether non-Hermiticity and EPs are a fundamental property of nature and if so, how nature utilizes them to gain competitive advantage have remained largely unanswered. Although biological systems feature many examples of non-reciprocal interactions with the potential to drive non-Hermitian dynamics, these are often theoretically overlooked and not experimentally investigated. Here, we demonstrate in an active matter composed of social animal *Caenorhabditis elegans* and bacteria, non-Hermitian dynamics, and the emergence of EPs owing to the nonreciprocal nature of oxygen sensing, nonequilibrium interfacial current, and bacterial consumption. We observed that when driven through the EP, the system collectively breaks parity-time (PT) symmetry leading to traveling waves and arrested phase separation. We further find that these features enable the collective ability to localize interfaces between broken and exact PT-phases. Remarkably, this ability provides a strong evolutionary advantage to animals living in soil. Altogether our results provide mechanistic insights into the detailed symmetries controlling the collective response of biological systems; answer a long-standing problem; and give an example of the EP-enabled dynamics in a biological system.

**Introduction**

Active matter systems consist of numerous energy-generating or consuming components, thus they are intrinsically out-of-equilibrium[1,2]. In these intricately coupled systems, independent interaction channels such as drift-diffusion processes, sensing, and social norms, naturally give rise to non-reciprocity[3-9] where the constituents of the system affect each other differently. This apparent violation of Newton's third law, which states action-reaction symmetry, is a common feature in active matter. This is especially true for biological systems which generally have multiple active and passive components (e.g., cells, dense swarming animals, growing tissues, and environmental matrix). Recent studies have shown that nonreciprocity (i.e., action-reaction symmetry breaking) plays a key role in collective behaviors in active matter and leads to many exotic phenomena, ranging from synchronization and flocking to dynamic pattern formation. Combining the framework of nonreciprocity and the interacting active material platforms promises to provide a powerful toolbox to dissect the complexities of living biological materials.

While nonreciprocity arises naturally in active material and biological systems, in many fields of science and engineering (e.g., optics, electronics, acoustics, etc.) one has to deliberately break time-reversal symmetry to induce nonreciprocal electromagnetic wave transmission or nonreciprocal interaction between the subcomponents of a system. This is often achieved through strong nonlinearities, space- and/or time-dependent modulation of constitutive material properties, and magneto-optical components. Most recently, nonreciprocal interactions and coupling have emerged as a resource for building highly sensitive sensors[10-12]; achieving unidirectional perfect absorbers; and suppressing and enhancing spontaneous emission, to name a few. Nonreciprocal interactions bring about non-Hermitian dynamics, which suggests the toolbox developed for studying non-Hermitian systems and the exotic features emerging from their EPs could be utilized in understanding and controlling complex active matter and biological systems.

Non-Hermiticity has its roots in quantum mechanics[13,14] and has been extensively studied in photonics[15], electronics, acoustics, optomechanics[16,17], superconducting qubits[18], trapped ions[19,20], single-spin systems[21], and in light-matter interactions[22]. Although in the majority of classical and quantum systems, non-Hermiticity and EPs emerge by judiciously controlling gain-loss balance as in active parity-time (PT) systems, dissipation- or loss-imbalance as in passive PT systems, and coupling strength among the couples, most recent studies have highlighted nonreciprocal coupling and interactions engineered using precisely located and controlled asymmetric scatterers or reflectors as a resource for non-Hermiticity[10,23,24]. In contrast to these artificially induced non-Hermiticity and PT symmetry, the majority of biological interactions are inherently nonreciprocal or asymmetric, and thus their dynamics can be modeled using effective Hamiltonian and dynamical matrix formalism widely used for non-Hermitian systems. Clearly, having nonreciprocal interactions, the active matter and biological systems should also have EPs and associated processes. Establishing this connection does not only allow studying complex biological systems using the well-known techniques utilized in non-Hermitian physics but it also will help answer the foundational question: How does the presence of EPs affect the dynamics of active matter and biological systems? Do presence of EPs and PT-symmetry breaking bring any advantage in biological systems? Despite significant progress in non-Hermitian physics and separately in active matter and biological systems, there is still a need for experimental platforms that bring together all these concepts to answer the above questions and reveal the potential biological implications of PT-symmetry, non-Hermiticity, and EP physics. Here we address this need and present experimental signatures of what happens to a biological system if it is driven through an EP between exact- and broken-PT phases, and how this affects the system's collective behavior.

In this study, we investigated the detailed nonequilibrium process together with the concept of nonreciprocity controlling the collective behavior of animals using *C. elegans* as a model organism. These animals come together and feed on bacteria lawns. This intriguing collective response is known as social feeding behavior[25-27]. The physics of this collective routine is remarkable because the mixture of active worms and passive bacteria forms a highly interacting multi-component condensate. This active mixture forms social groups during feeding. Previous genetic studies have linked this feeding behavior to oxygen sensing[26,28], which promotes tracking low oxygen levels to locate bacteria as food. More interestingly, during their domestication process[29] as model organisms in the lab, the natural isolates of these social worms acquired several genetic mutations that significantly altered their oxygen preferences. As a result, social strains became solitary in the lab. Taken together, this collective behavior and the variability of their social response provide a valuable experimental system to study the detailed non-equilibrium dynamics of this interacting active matter system.

We found that all theoretically predicted non-Hermitian features, some of which are already observed in non-biological systems, including arrested coarsening, transitions between traveling and standing waves, and edge localization and delocalization emerge in these social animal groups. Using the approach learned from non-Hermitian physics in a non-equilibrium regime, our findings shed light on understanding the complex behaviors of biological systems including their evolutionary significance.

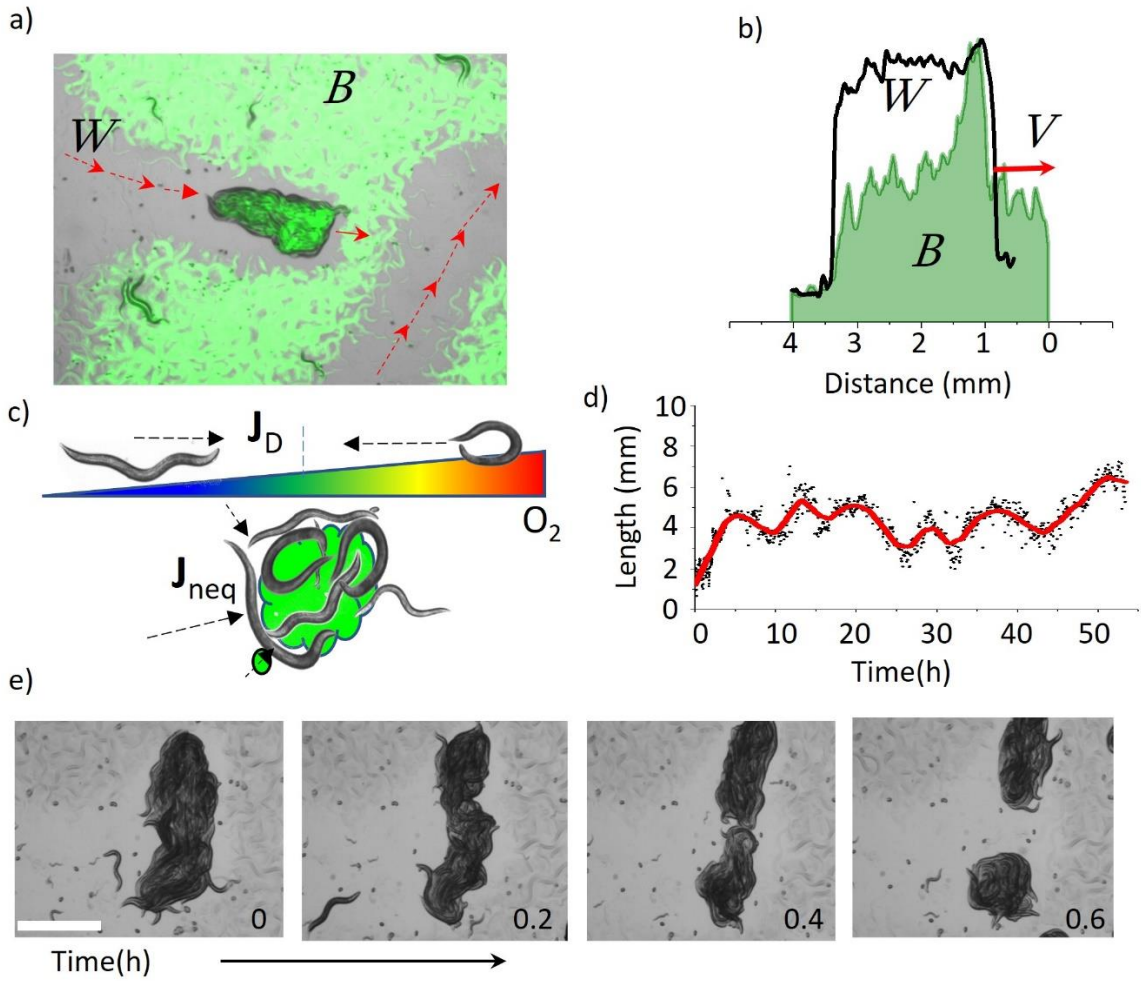

*Figure 1 Emergence of a traveling wave state in an active worms-bacteria mixture. a) The image of a worm aggregate overlaid with a GFP-labeled bacterial field, where a group of worms is swarming together. b) Densities of the worm W and bacteria B fields, revealing the asymmetric profile during the traveling state. V represents the forward velocity of the animal group. c) Schematic representation of how active worm aggregates concentrate bacteria by coming together on a bacteria lawn and reducing oxygen levels. Conversely, aggregated worms can drift bacteria across the interface. This process could be represented by two material currents $J_D$ and $J_{neq}$. d) The size of the worm aggregate as a function of time indicates the arrested coarsening during the traveling state. e) Splitting of worm aggregate into smaller parts limiting the growth of the aggregate size. Scale bar, 1mm.*

## Results

**Experimental observation of the traveling state of coupled worms and bacteria mixture**

To study the physics of social feeding behavior involving bacteria, we used time-lapse microscopy to observe worms (***W***) and bacteria (***B***) densities together (Fig. 1). We used the Green Fluorescent protein (GFP) to label the bacteria so that we could observe the worms and bacteria separately. Over the course of several days of imaging starting from a single worm, we observed the formation of small groups of animals traveling on the bacteria lawn (Figure 1a, Supplementary Video1-2). During this traveling state, the worms and bacterial fields showed various configurations, including colocalization, delocalization, and asymmetric density distributions (Figure 1b). Of particular interest is the asymmetric density profile, which arises when the animals are placed on a flat and uniform bacterial lawn. In this scenario, the swarming animals can spontaneously develop a bacterial gradient and move towards regions with higher bacterial density.

In a previous study[30], we observed that when worms aggregate, they can cause bacteria to concentrate within the aggregates, leading to the formation of complex dynamical patterns. Initially, we attributed this to the low oxygen taxis behavior of the worms, which caused the bacteria to co-locate due to the sponge-like structures in the animal groups (Figure 1c). However, it turned out that the process was more exotic, and was purely driven by their activity. This colocalization process was the first dynamical feature that prompted us to investigate the active matter nature of the worm-bacteria mixture, which was likely responsible for the observed phenomena. In this study, we further focused on several other experimental observations that suggested that this process was particularly triggered by a nonequilibrium and hence non-Hermitian process in the animal groups.

Another signature of non-equilibrium behavior we noticed in this animal model is the arrested size of the active worm aggregates (Figure 1d). Animal groups can merge together, but if they form larger groups, they later split into smaller parts (Figure 1e, Supplementary Video 3). The arrested coarsening dynamics of the active system is of special importance because the process limiting the universal coarsening event requires critical dynamics. The most parsimonious hypothesis explaining this observation is that the interface of the worm aggregates generates a nonequilibrium bacterial current ($\mathbf{J_{neq}}$) that could limit the coarsening process[31].

Finally, we observed that the worm-bacteria condensates were highly vulnerable to local depletion of bacterial densities, leading to the formation of bubbles (Supplementary Video 4). These observations share some similarities with the recently developed active model B+[32], also known as bubbly phase separation, which is based on a single active component. However, it is important to note that our system differs from one-component active platforms due to the coupling of the worm and bacterial fields. Despite this difference, the macroscopic dynamics of hole formation in the worm-bacteria mixture were found to be similar. These three experimental findings support the idea that the worm-bacteria active system exhibits special nonequilibrium dynamics that control their collective behaviors, but they do not tell whether EPs exist in this model system and what roles EP, if exist, and PT-symmetry breaking play in the observed behaviors.

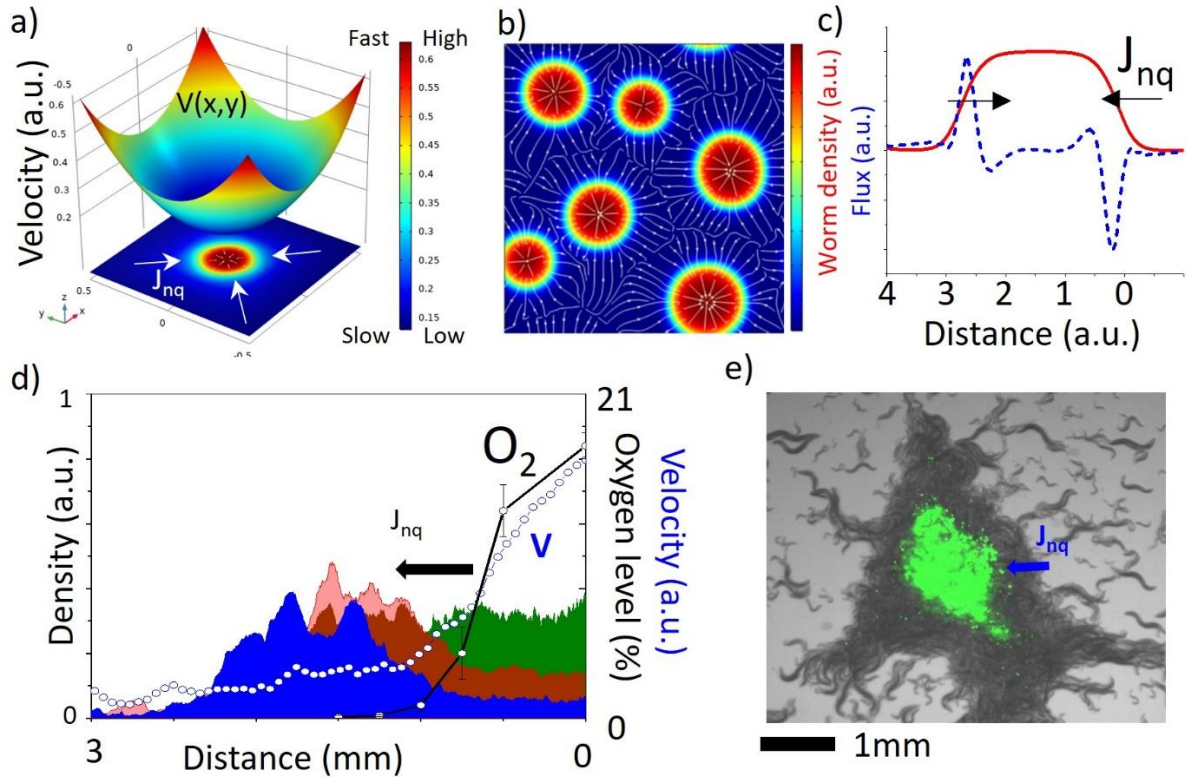

*Figure 2: Spatial activity gradient generates non-equilibrium interfacial current on the passive component. a) Numerical simulation demonstrating the drift process driven by a velocity gradient in space. b) Simulation result showing the aggregating active worms generate a velocity gradient across the interface, resulting in a drift current $J_{nq}$. c) The cross-sectional profile of worm density and the corresponding nonequilibrium drift current. d) Experimental results demonstrating the interfacial drift current of passive beads from the oxygen-lacking edge to the oxygen-depleted center of the group. Aggregating worms and self-consumption decrease oxygen concentration and generate a spatial activity gradient. Green (initial) to blue (final) colors indicate the time and bead distribution towards low oxygen regions where the animals are slow. e) A typical image of a worm aggregate overlaid with concentrated passive green fluorescent beads at the center of the group. Note that worm and bead densities are colocalized. Oxygen and velocity profiles are given in Supplementary Figure1.*

**Spatial activity gradient generates non-equilibrium interfacial current**

To further assess the nonequilibrium process at the interface of the worm-bacteria mixture, we performed numerical simulations. We found that the speed of the worms was strongly dependent on the oxygen concentration, while the bacteria were passive constituents of the system that exhibited only small fluctuations. Due to hydrodynamic coupling with the bacteria, the worms were capable of inducing strong activity at large scales, leading to bacterial drift and spatial variations in the speed at the interface[33-35] (Supplementary Video5).

It is important to note that this process involves the coupling of large active worms (~1mm) and small passive bacteria (~5μm), which is critical due to the different types of drift mechanisms that originate from the size difference between the interacting components. First, under a spatial activity gradient, small active matter can apply active phoretic pressure to larger passive components, as seen in the centering of the nucleus in a cell, where the activity gradient around the cortex pushes large particles towards the center[36,37]. Additionally, at the cellular membrane, the dynamic coupling between large cargo proteins and the active MinB system can also be described as an example of phoretic active pressure[38]. From this perspective, our system exhibits different dynamic processes. The second type of drift originates from the spatial activity dictated by large active particles. It is worth noting that the response of particles to activity gradients has received significant scientific attention and has been theoretically well-studied[33,39]. Our system should be considered in this class.

To better understand the dynamics of the interaction between large active worms and small passive bacteria at interfaces, we simulated bacterial diffusion, where the speed of the passive bacteria is dictated by the active worms. The worms can also self-aggregate due to local oxygen depletion, which we modeled by using the same principle of motility-induced phase separation, resulting in negative effective diffusion[40,41] ($D_{eff} < 0$, Supplementary Note section 1, Supplementary Video 6). This self-aggregation condition can be seen as the first of two instabilities required for non-reciprocal phase transitions. As expected, the activity gradient at the interface generates a drift current, causing the passive particles to be pumped toward the center of the aggregates (Figure 2a-c, Supplementary Video 6). We observed the same process even when bacteria were replaced with polymer beads, further confirming the nonequilibrium nature of the process (Figure 2d-e, Supplementary Figure 1). The dense worm aggregates depleted oxygen, leading to the formation of oxygen and velocity gradients at the interface, which pumps the passive beads toward the center. We observed similar nonequilibrium interface currents and co-localization of active and passive components at a larger scale (Figure 2e). Based on these experimental findings and numerical results, the dynamics of the entire process can be reduced to a flux term at the interface, which we modeled using $J_{neq} = \zeta B \nabla W$ (Supplementary Note section 2).

**Emergence of exceptional points and traveling state in worm-bacteria mixture**

To gain more intuition about the coupled dynamical system between worms and bacteria and study how EPs emerge in this dynamics, we formulated a set of coupled drift-diffusion equations for two conserved density fields, **W** and **B**. The time evolution of the worm density is influenced by both oxygen-dependent motility and the interaction between worms and bacteria. These dynamics can be expressed mathematically as follows:

$$\frac{\partial W}{\partial t} = \nabla \cdot [D_{eff} \nabla W + \beta W \nabla B] - \gamma_W \nabla^4 W$$

$$\frac{\partial B}{\partial t} = \nabla \cdot [D_B \nabla B - J_{neq}] - \lambda W B - \gamma_B \nabla^4 B$$

where $D_{eff}$ represents the motility-dependent dispersion of worms which is negative and promotes self-aggregation, $D_B$ is the diffusion coefficient of bacteria induced by active worms, and $\beta = \frac{v}{\tau}\frac{\partial v}{\partial B}$ is the aerotactic coupling coefficient indicating the strength of the animal response to the bacteria-dependent oxygen gradient. Note that $\lambda$ is the bacterial consumption by the worm and $\gamma$ is the phenomenological surface tension parameter. $v$ and $\tau$ are the velocity and reversal rates of the animals.

As a new cross-coupling term we added non-equilibrium flux ($\vec{J_{neq}} = \zeta B \nabla W$) which is the first term breaking the reciprocity between worms and bacteria. When we linearize the system around equilibrium points, the dynamical matrix can be simplified to the version given below, which shares the common form of the coupled dynamical system widely used in non-Hermitian physics (Supplementary Note section 3)

$$\frac{\partial}{\partial t}\begin{bmatrix}\rho_W\\\rho_B\end{bmatrix} = \begin{bmatrix}D_{WW} & D_{BW}\\D_{WB} & D_{BB}\end{bmatrix}\begin{bmatrix}\rho_W\\\rho_B\end{bmatrix}$$

$D_{BW}$ is the cross-diffusion term and represents the chemotactic drift of the worms across the bacterial gradient. This term is controlled by the oxygen concentration which is defined by the local bacterial density. Further, the other cross-diffusion term $D_{WB}$, absorbs non-equilibrium interface flux ($\vec{J_{neq}}$) promoting colocalization and also the consumption rate of bacteria ($\lambda$) by the worms which acts as a delocalization term in the system. The schematic representation of these interactions between worms and bacteria is given in Figure 3a. Eigenvalues $\sigma_\pm = (D_{WW} + D_{BB})/2 \pm \sqrt{\xi}/2$ where $\xi = (D_{WW} - D_{BB})^2 + 4D_{WB}D_{BW}$ reveals the pseudo-Hermitian characteristic of the system (Supplementary Note section 3,4) with the emergence of an EP at $\xi =$ 0 where both the eigenvalues and the associated eigenvectors coalesce. EP divides the parameter space into three: i) $\xi = 0$ where eigenvalues are degenerate (i.e., critically damped harmonic oscillator); ii) ) $\xi < 0$ eigenvalues become complex conjugate pairs (i.e., underdamped harmonic oscillator) and the system oscillates; and iii) ) $\xi > 0$ where two distinct eigenvalues emerge and the system approaches the equilibrium position without any oscillation (i.e., overdamped harmonic oscillator). We note that the emergence of EP is the second instability indicating the Parity-Time (PT) symmetry-breaking conditions (Figure 3b, Supplementary Note section 4,5 ). The eigenvalues of the system be simply controlled by two critical external parameters; consumption rate $\lambda$ of bacteria by worms which can balance the bacterial pumping into the worm aggregates and also ambient oxygen level that controls the activity and the sensitivity of the worms. The corresponding phase space has three major domains, uniform densities, static pattern forming, and traveling state regions (Figure 3b top). To further test the theoretical predictions, we numerically solved the coupled system in two dimensions (2D) (Supplementary Video 7). We observed that static colocalized (aligned), delocalized (antialigned) patterns, and traveling (chiral) states emerged during the simulation (Figure 3c). We then repeated the simulations by implementing local initial noise to trigger the first aggregation instability and observed the dynamics of localized individual groups traveling toward bacteria-available regions (Figure 3d). The system spontaneously breaks spatial symmetry and develops a self-generated bacterial gradient profile (Figure 3e). This broken symmetry further guides animal groups into spontaneously picked outward directions (Supplementary Video 8,9). When we plot the density profiles of worms (W) and bacteria (B) the broken symmetry and stable phase difference between these fields become more evident (Supplementary Note section 4). This phase difference also indicates the chiral state formed by the worm bacterial fields. Simply, worms are chasing the self-generated bacterial gradient profile.

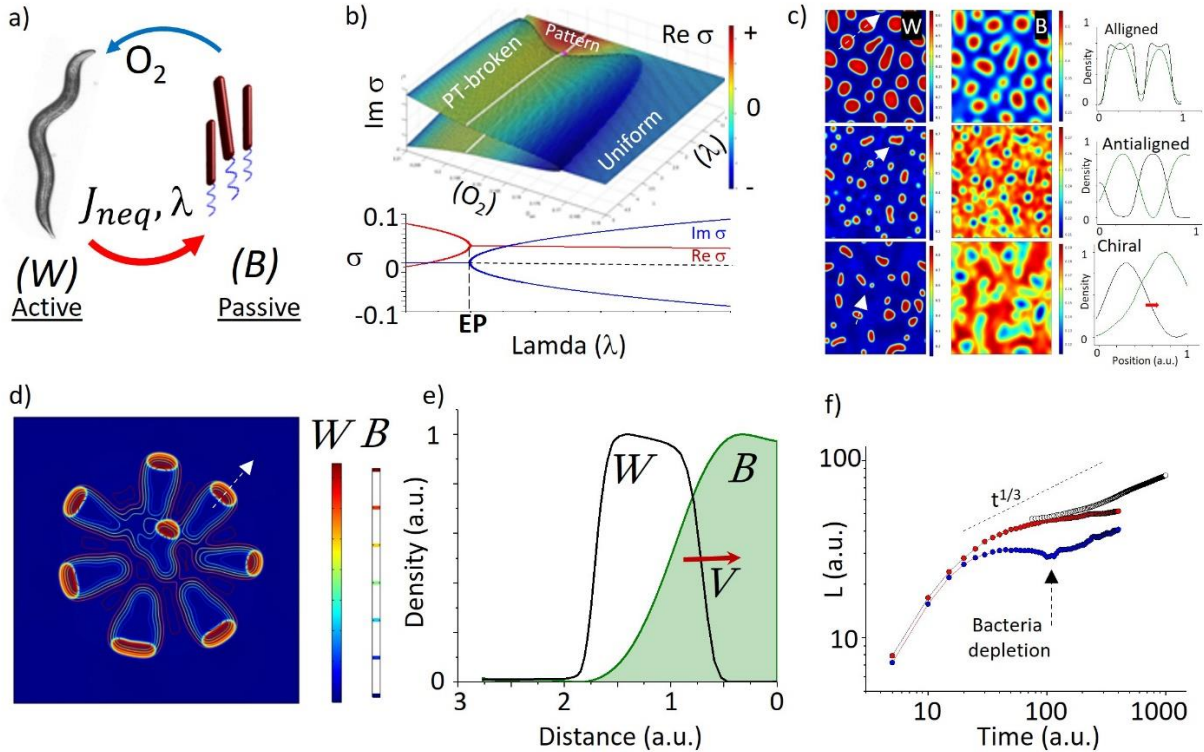

*Figure 3: Emergence of exceptional points (EPs) in nonreciprocally interacting worm and bacterial active mixture. a) Schematic representation of a two-component active matter system consisting of worms (W) and bacteria (B). Non-equilibrium fluxes control the time evolution of the system by promoting colocalization and bacterial consumption λ drives the delocalization process of worm and bacterial fields. Due to aerotaxis, the depletion of oxygen by the bacteria controls the worm's activity. b) Numerical simulation of the phase diagram of the coupled active matter mixture indicating three different domains. The system dynamics depends on the critical coupling parameter λ and the ambient oxygen level. Above a critical level, the eigenvalues of the system develop complex conjugate pairs (bottom). The emergence of the complex conjugate pairs drives the traveling state of the patterns (PT broken region). c) Worm and bacterial fields show aligned (colocalized) and antialigned (delocalized) profiles before forming the traveling state (chiral). Intensities are measured across the arrows. d) Numerical simulation of the worm aggregates traveling across uniform bacterial density (contour). e) Simulation results of worm and bacterial densities indicating the asymmetric density profile above EP indicating the PT-symmetry breaking between worm bacteria fields. f) The coarsening process of the worm aggregation is arrested (Blue). Numerical simulation indicates the separate impact of bacterial consumption (red) and interface current on the coarsening dynamics. Black dots represent the result of the universal coarsening process without bacterial current and consumption.*

As a final stage of the numerical simulations, we analyzed the coarsening dynamics of the worm bacteria mixture under various conditions. We first tested the characteristic size of the groups above the EP where the aggregates are moving. As expected in this regime the coarsening is arrested (Figure 3f). This arrested state is easily understood by considering the bacterial flux around the periphery of the groups. As groups grow in size the interface has to pump more bacteria to maintain the low oxygen level around the favorable range. Above the critical size; however, bacteria get depleted due to high consumption and oxygen level increases which further promotes high motility. The increase in the motility increases the diffusion and finally breaks the groups into small parts as we observed in our experiments. In the case of worm condensate (Supplementary Video 4), high consumption of bacteria could locally trigger the oxygen increase and the final system could drive the formation of holes. This point is the similarity between our system and the active Model B+. The susceptibility of the system against local noise is mainly triggered by the imbalance between inward bacterial pumping and oxygen increase. We also tested the coarsening scenario without bacterial consumption ($\lambda = 0$) while keeping nonequilibrium pumping on ($\vec{J_{neq}}$). Interestingly the system shows different scaling. We hypothesize that this suppression is controlled by the interface current of bacteria in the aggregates. Finally, we noticed that removing all the cross-coupling terms between worms and bacteria gave rise to the formation of regular motility-induced phase separation with a universal coarsening profile $\sim t^{1/3}$ (Figure 3f). All these numerical results support that the active worm and bacteria mixture has non-Hermitian features and shows all expected dynamical properties including EPs, traveling state, and arrested coarsening process. Nonequilibrium interface current and bacterial consumption together spontaneously breaks PT symmetry leading to the emergence of a collective traveling state of the worm aggregates across uniform bacterial density. The similarities between theoretical expectations and the experimental results are remarkable.

**Edge localization and evolutionary significance**

One of the interesting observations in topological physics is the formation of localized edge states at the interface of two topologically different domains[42-46]. Such edge states could also emerge in non-Hermitian systems. Recent theoretical studies have elaborated on the possibility of such dynamics in various biological networks[44-46]. In this study, we aim to bring a different perspective to this concept. We realized that interfaces are very common environmental features in nature, such as water-soil interfaces in lake sediments or air-soil interfaces on the ground surface of forests. These interfaces are the main battlegrounds between competing animal species. Different regions of these interfaces may have separate challenges or predators, and staying locally around them could provide specific advantages. Similarly, soil nematodes, *C. elegans,* live in wet soil where the fluids are decomposing. Understanding the collective response of the worms at these interfaces (Figure 4a) may link the physics of localized edge states and their potential biological implications.

In order to elucidate topology in interacting biological systems, we studied the dynamics of worms, specifically their behavior around interfaces representing different symmetries in the phase space. We selected the interface between exact and broken PT domains which were obtained by partially blocking oxygen penetration from the air. The numerical results showed the localization of animals along the interface (Figure 4b and Supplementary Video10). In our experiment, we placed a cover glass on a bacteria lawn with swarming animals and observed their collective responses. Our findings indicate that interfaces can lead to the emergence of localization at the sharp interface where oxygen concentration quickly shifts from high (normoxic) to low (anoxic) levels. Neither side of the interface is favorable for single worms; the open region has high oxygen levels (i.e., broken PT phase) and is open to predators, while the closed region has very low oxygen levels (i.e., exact PT phase) due to the presence of bacteria and blocked oxygen penetration. However, at the interface, animals can spot favorable conditions and localize by slowing down (Figure 4c, Supplementary Video11). This raises the question of whether localization provides an evolutionary advantage. To test this hypothesis, we designed a new experiment to mimic the granular structure of the soil, the natural habitat of *C. elegans*. We used gel-based beads (Sephadex 50) soaked in bacterial suspension (as shown in Figure 4d), which has low oxygen levels but provides more interfaces for worms to stay around. To test the worms' preference, we extended the bacterial lawn to an open region without interfaces. We found that the majority of social worms (npr-1) quickly found and preferred staying in the granular region when placed away from it, while solitary strains (N2) showed a different response, spreading around without a clear preference for this region (Figure 4e, Supplementary Video12). Based on these observations we can conclude that the natural isolate of *C. elegans,* aka social strain*,* prefers staying at the interfaces where they can get sufficient oxygen while keeping themselves from open regions. When they get into the open region they come together and collectively form groups by decreasing internal oxygen levels. Interestingly solitary strain N2 which evolved in the lab on a flat surface lost this collective response and interface tracking ability.

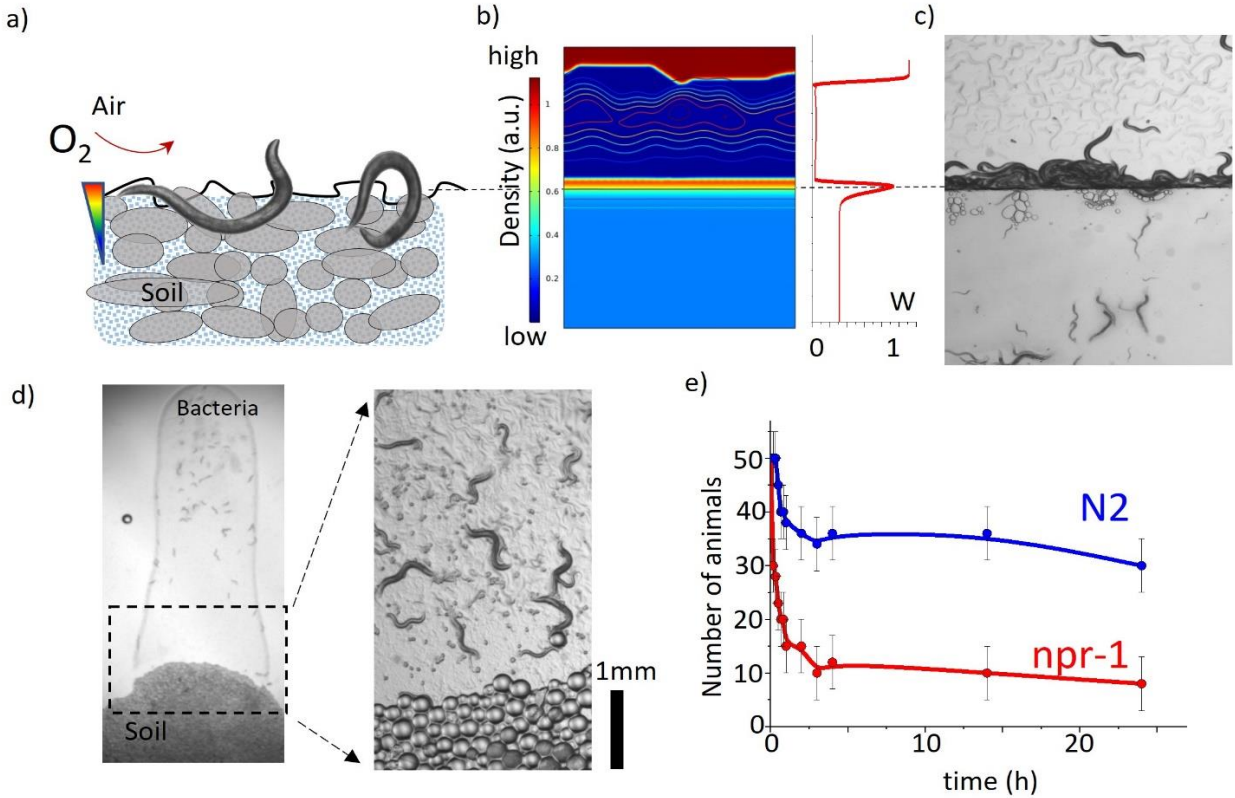

*Figure 4: Evolutionary significance of edge localization of social animals at the interface between different domains. a) Schematic representation of an air-soil interface corresponding to uniform and traveling states of the phase diagram. b) Numerical results of the coupled worm-bacteria system localizing around the edge of the air-soil interface are shown. c) Experimental result of the localized edge state at the interface generated by blocking oxygen penetration. d) Image of the experimental platform used to test the ability to localize around the edge. Polymer beads are used to generate a multilayered structure to mimic the soil-air interface. e) Social animals prefer staying in the granular region where they can hide and stay around the edges. However, solitary animals spread around and do not show any edge preference.*

## Discussion

Nonreciprocity, a characteristic phenomenon observed in interacting biological entities—from bacteria to birds in flight—is inevitable in nature. Influenced by various factors like social interactions, restricted perceptual abilities, or hydrodynamics, this nonreciprocity plays a crucial role and serves as a bridge, linking the dynamics of biological systems to non-Hermitian physics, a field with significant applications in areas such as optics, electronics, acoustics, and quantum mechanics. Such a combination provides a powerful toolbox for probing the complex dynamics of active matter, offering fresh insights into the intricacy of biological systems.

Non-Hermitian, including PT-symmetric, toolbox, and features offer a beneficial macroscopic metric to understand a system's response. In our study, we noted that PT symmetry breaking manifests as an arrested travel state of animal groups. This observation is critical since pattern-forming systems typically generate standing wave patterns with a universal coarsening response[47]. Traveling wave patterns akin to these have been studied in various physical systems, like viscous fingering[48], which necessitates two successive instabilities to produce traveling waves. The primary difference in active biological systems is their self-aggregating behavior, corresponding to the first instability. The role of nonreciprocity becomes particularly significant for the second instability, which spontaneously breaks PT symmetry and provides collective motility.

Moreover, we discovered that interface localization is a crucial difference between social and solitary animals. This response is indeed vital in nature but not in the lab due to the absence of the predators, and this finding illuminates why the solitary strain lost this ability during the lab domestication process. The most captivating aspect of physical systems is edge localization at the interface of two topologically different systems (i.e., the interface of systems with trivial and nontrivial topologies). This characteristic, emerging from topological constraints, allows for unidirectional propagation along the interface, showing resilience against external disturbances. Our findings also introduce new questions, like the chirality of localized animals at the interfaces having varying symmetries. We hypothesize that fluctuations of the localized state around the interface should propagate unidirectionally. Future investigations are necessary to explore these intriguing collective behaviors of animals.

In conclusion, different forms of nonreciprocity can be simultaneously observed in large microbial populations. Our findings hold potential relevance for comprehending the complex dynamics of these populations, from gut microbiota to ecological systems. We further speculate that the principles derived from non-Hermitian physics could illuminate the understanding of interactive biological systems.